\documentclass[preprint,showpacs,preprintnumbers,superscriptaddress,amsmath,amssymb]{revtex4}
\usepackage{amssymb}
\usepackage{graphicx}
\usepackage{float}
\begin{document}
\title{The production of unknown neutron-rich isotopes in $^{238}$U+$^{238}$U collisions at near-barrier energy}
\author{Kai Zhao}
\email{zhaokai@ciae.ac.cn} \affiliation{Department of Nuclear Physics, China Institute of Atomic Energy, P.O. Box 275(10), Beijing 102413, People's Republic of China}
\author{Zhuxia Li}
\email{lizwux@ciae.ac.cn} \affiliation{Department of Nuclear Physics, China Institute of Atomic Energy, P.O. Box 275(10), Beijing 102413, People's Republic of China}
\author{Yingxun Zhang}
\affiliation{Department of Nuclear Physics, China Institute of Atomic Energy, P.O. Box 275(10), Beijing 102413, People's Republic of China}
\author{Ning Wang}
\affiliation{Department of Physics, Guangxi Normal University, Guilin 541004, People's Republic of China}
\author{Qingfeng Li}
\affiliation{School of Science, Huzhou University, Huzhou 313000, People's Republic of China}
\author{Caiwan Shen}
\affiliation{School of Science, Huzhou University, Huzhou 313000, People's Republic of China}
\author{Yongjia Wang}
\affiliation{School of Science, Huzhou University, Huzhou 313000, People's Republic of China}
\author{Xizhen Wu}
\affiliation{Department of Nuclear Physics, China Institute of Atomic Energy, P.O. Box 275(10), Beijing 102413, People's Republic of China}

\date{\today}

\begin{abstract}
The production cross sections for primary and residual fragments with charge number from $Z$=70 to 120 produced in the collision of $^{238}$U+$^{238}$U at 7.0 MeV/nucleon
are calculated by the improved quantum molecular dynamics (ImQMD) model incorporated with the statistical evaporation model (HIVAP code).
The calculation results predict that about sixty unknown neutron-rich isotopes from element Ra ($Z$=88) to Db ($Z$=105) can be produced
with the production cross sections above the lower bound of $10^{-8}$ mb in this reaction.
And almost all of unknown neutron-rich isotopes are emitted at the laboratory angles $\theta_{lab}\leq$ 60$^\circ$.
Two cases, i.e. the production of the unknown uranium isotopes with $A\geq$ 244 and that of rutherfordium with $A\geq$ 269 are investigated for understanding the production mechanism of unknown neutron-rich isotopes. It is found that for the former case the collision time between two uranium nuclei is shorter and the primary fragments producing the residues have smaller excitation energies of $\leq$ 30 MeV and the outgoing angles of those residues cover a range of 30$^\circ$-60$^\circ$. For the later case, the longer collision time is needed for a large number of nucleons being transferred and thus it results in the higher excitation energies and smaller outgoing angles of primary fragments, and eventually results in a very small production cross section for the residues of Rf with $A\geq$ 269 which have a small interval of outgoing angles of $\theta_{lab}$=40$^\circ$-50$^\circ$.

\end{abstract}


\maketitle

\section{Introduction}

The production of unknown neutron-rich nuclei, especially for unexplored superheavy nuclei and isotopes near $r$-process, in fusion, fission, fragmentation process and multinucleon transfer reactions has been of experimental and theoretical interest. A lot of more neutron-rich isotopes below Ra in fragmentation process were produced in recent years\cite{Kurt4616}. But for the new nuclei in the 'northeast' area of the nuclear map, it is difficult to be reached in the fission reactions and fragmentation processes widely used nowadays. Due to the 'curvature' of the stability line, it is also difficult for reaching these new more neutron-rich nuclei in fusion reactions with stable projectiles because of the lack of neutron number. The revival interest of multinucleon transfer between actinide nuclei at low-energy collisions, such as two $^{238}$U, has arisen. This type of reaction provides us with an alternative way to produce more neutron-rich actinide and transactinide isotopes through multinucleon transfer.

During later 1970s and early 1980s, the uranium beam available at GSI was used to investigate the gross features of the products of reactions $^{238}$U+$^{238}$U and $^{238}$U+$^{248}$Cm associated with the charge distribution and cross sections for heavy actinide isotopes\cite{Hild1065,Scha1978,Esse0265,Frei0171,Sch1982,Mood1315}. The experimental data of the actinide and complementary products in the reaction $^{238}$U+$^{238}$U were reexamined in 2013\cite{Krat4615}.
At GANIL, the experiment on the collision of
$^{238}$U+$^{238}$U at energies between 6.09 MeV/nucleon and 7.35 MeV/nucleon
was performed, and the dependence of production yield of products on the beam energy and on the angle of detection were measured\cite{Gola2235}.
However, new neutron-rich nuclei have not been experimentally reported in the reaction of $^{238}$U+$^{238}$U up to now.

The production of unknown neutron-rich isotopes was predicted in low-energy dissipative collisions of $^{238}$U+$^{248}$Cm through multinucleon transfer
based on multidimensional Langevin equations\cite{Zagr0366,Zagr4608}.
The semiclassical model GRAZING with considering the competition between neutron emission and fission showed the production of a few unknown neutron-rich isotopes with $Z$=92 to 94 in $^{238}$U+$^{238}$U at entering energy $E_{lab}$=2059 MeV\cite{Yane4608}.
Due to a large number of degrees of freedom, such as deformations of two nuclei, neck formation, nucleon transfer, nucleon emission and different types of separation of the transient composite system being involved in the reaction, it is more suitable to apply
a microscopic dynamical model to investigate the reaction mechanism and the production of unknown isotopes.
TDHF approach was used to analyze the role of nuclear deformation on collision time and on nucleon transfer in central collisions of $^{238}$U+$^{238}$U,
but the production of residual fragments was not calculated yet\cite{Gola2701,Davi4613}. Microscopic transport model such as QMD type models were also applied to study the low energy reactions of heavy nuclei systems such as $^{197}$Au+$^{197}$Au, $^{238}$U+$^{238}$U
and $^{232}$Th+$^{250}$Cf\cite{Maru0191,Wang2014a,Wang2014b,Wen2013,Wang2619,Tian4603,Zhao2009,Zhao2013}.
By ImQMD model incorporated with the statistical evaporation model (HIVAP code)\cite{Wrei0154,Shen1602}, the mass distribution
of products in $^{238}$U+$^{238}$U at 7.0MeV/nucleon was calculated and
generally in consistence with the experiment measurement of
GANIL\cite{Zhao2009}. The calculated
isotope distributions of the residual fragments and the most probable mass number of fragments were generally
in agreement with experimental data of GSI, and the production mechanism of neutron-rich residual fragments was studied\cite{Zhao2015}.
In this work, we will further investigate the production of the primary and residual fragments with charge number from $Z$=70 to 120 produced in reaction of $^{238}$U+$^{238}$U at 7.0 MeV/nucleon.

The structure of this paper is as follows. In Sec. II, the
framework of the ImQMD model is briefly introduced. In Sec. III,
the production cross sections for primary and residual
fragments are calculated, and the production of unknown neutron-rich isotopes in the reactions of $^{238}$U+$^{238}$U will be discussed.
Further, the microscopic mechanism of producing these isotopes is carefully analyzed. Finally, a brief summary
is given in Sec. IV.

\section{Theoretical Model}

As in the original
QMD model\cite{Aich0233,Aich0034,Hart0303}, each nucleon is represented by a coherent
state of a Gaussian wave packet in the ImQMD model.
The time evolution of the coordinate and momentum for each nucleon in
the mean field part is determined by Hamiltonian equations.
The Hamiltonian includes kinetic energy, nuclear potential energy
and the Coulomb energy. The nuclear potential energy is an
integration of the Skyrme type potential energy density
functional, which reads
\begin{eqnarray}
V_{loc}=&&\frac{\alpha}{2}\frac{\rho^{2}}{\rho_{0}}+\frac{\beta}{\gamma+1}
\frac{\rho^{\gamma+1}}{\rho _{0}^{\gamma}}+\frac{g_{0}}{2\rho _{0}}\left(\nabla\rho\right)^{2}\nonumber\\
&&+\frac{c_{s}}{2\rho_{0}}[\rho^{2}-\kappa_{s}(\nabla\rho)^{2}]\delta^{2}+g_{\tau}\frac{\rho^{\eta+1}}{\rho_{0}^{\eta}},
\end{eqnarray}
where $\rho=\rho_{n}+\rho_{p}$ is the nucleon
density and
$\delta=(\rho_{n}-\rho_{p})/(\rho_{n}+\rho_{p})$ is the isospin asymmetry.  $\rho_{n}$,
$\rho_{p}$ are neutron and proton density, respectively.

The Coulomb energy is written as a sum of the direct and the
exchange contribution:
\begin{equation}
U_{Coul}=\frac{1}{2}\int \int \rho _{p}(\mathbf{r})\frac{e^{2}}{|\mathbf{r-r}%
^{\prime }|}\rho _{p}(\mathbf{r}^{\prime
})d\mathbf{r}d\mathbf{r}^{\prime }-e^{2}\frac{3}{4}\left(
\frac{3}{\pi }\right) ^{1/3}\int \rho _{p}^{4/3}d\mathbf{R}.
\end{equation}

In the collision part, the phase space occupation constraint for single particle proposed by Papa \textit{et al.}\cite{Papa4612} is applied in each time evolution step. The isospin-dependent in-medium
nucleon-nucleon scattering cross sections are applied. The Pauli-blocking effect is treated as the same as in reference\cite{Qing4612}, which is obtained according to the the Uehling-Uhlenbeck factor. The model parameters
as those used in Ref.\cite{Zhao2013} are listed in TABLE
\ref{parameter}. More detailed description of ImQMD model and its applications can
be found in Refs.\cite{Wang4608,Wang2004,Wang2619,Zhao2009,Zhao2013}.
\begin{table}[htbp]
\vspace{10mm}
\begin{center}
\caption{\label{parameter}the model parameters}
\begin{tabular}{ccccccccc}
 \hline
  $\alpha$(MeV) & $\beta$(MeV) & $\gamma$ & $g_0$(MeV$fm^{2}$) & $g_{\tau}$(MeV) & $\eta$ & $c_{s}$(MeV) & $\kappa_{s}$($fm^{2}$) & $\rho_0$($fm^{-3}$)\\
  \hline
  -356 & 303 & 7/6 & 7.0 & 12.5 & 2/3 & 32 & 0.08 & 0.165\\
 \hline
\end{tabular}
\end{center}
\end{table}

In this work, the binding energy per nucleon and
deformation of $^{238}$U are taken as $E_{gs}$= 7.37 MeV,
$\beta_2$=0.215 and $\beta_4$=0.093 given by Ref.\cite{Pmol0185}.
For low-energy collision of $^{238}$U+$^{238}$U,
the initial condition of reaction, such as the properties of
projectile and target nuclei, is of vital importance for the microscopic transport model.
We check the binding energy, the
root-mean-square radius and the deformation of the initial nuclei,
as well as their time evolution carefully. Only those initially selected nuclei with no spurious particle
emission and their properties, such as the binding energy,
root-mean-square radius and deformation being stable within 1000fm/c are adopted. The orientations of the initial
uranium nuclei in all events are sampled randomly with an equal
probability.
In the ImQMD model, the time evolution of the reaction for each
event at different impact
parameters can be tracked. Both the formation time and the
reseparation time of the transient composite system of $^{238}$U+$^{238}$U can be recognized in the simulations\cite{Zhao2015}. The charge number $Z$, mass number $A$ and the excitation
energy $E^*$ of each fragment formed in each event can also be determined. The cross section for producing the
primary fragment with $Z$, $A$ and $E^*$ is then calculated by
\begin{eqnarray}
\sigma(Z,A,E^*)& &=\int_0^{b_{max}}2\pi{}bdb\frac{N_{frag}(Z,A,b,E^*)}{N_{tot}(b)}\nonumber\\& &
=\sum_{b=0}^{b_{max}}2\pi{}b\Delta{}b\frac{N_{frag}(Z,A,b,E^*)}{N_{tot}(b)}.
\label{primary-cross-section}
\end{eqnarray}
Here $b$ is the impact parameter, $N_{frag}(Z,A,b,E^*)$ is the number
of events in which a fragment $(Z,A,E^*)$ is formed at a given
impact parameter $b$. The excitation energy $E^*$ of the fragment
with charge number $Z$ and mass number $A$ is obtained by
subtracting the corresponding ground-state energy\cite{Pmol0185}
from the total energy of the excited fragment in its rest frame.
$N_{tot}(b)$ is the total event number at a given impact parameter
$b$. The outgoing angle of each primary fragment can be
obtained from its momentum. In this work, the maximum impact parameter is
taken to be $b_{max}$=15 fm, and the impact parameter step is
$\Delta{b}$=0.15 fm. The initial distance between the centers of
mass of projectile and target is taken to be 40 fm. 100,000 events
for each impact parameter are simulated in this work.

At 1000fm/c after the re-separation of the composite system, the ImQMD simulation is terminated and the primary fragments are recognized at this time as that did in Ref.\cite{Zhao2015}.
Then the de-excitation process, including the evaporation of $\gamma$, $n$, $p$ and $\alpha$ particle and fission, for each excited primary fragment is performed by using the statistical evaporation model (HIVAP code)\cite{Wrei0154,Shen1602}.
In the HIVAP code, the survival probability of the fragment with charge number $Z$, mass number $A$ and excitation energy $E^*$ are calculated by branching ratios expressed by relative partial decay widths for all possible decay modes, $\Gamma_{i}(Z,A,E^*)/\Gamma_{tot}(Z,A,E^*)$, where $\Gamma_{tot}(Z,A,E^*)=\sum_i\Gamma_{i}(Z,A,E^*)$, and
$i$=$\gamma$, $n$, $p$, $\alpha$, and fission.

\section{Results and Discussion}

\begin{figure}[hbtp]
\includegraphics[angle=0,scale=0.8]{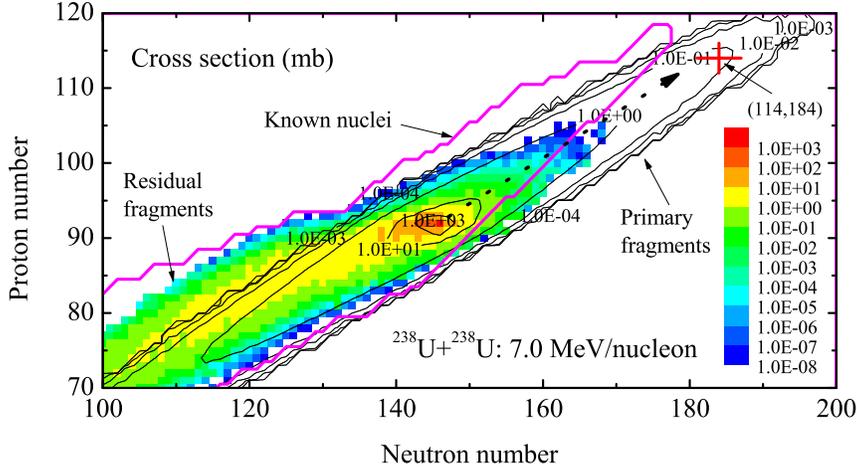}
\caption{(Color online) The landscape of the cross sections for primary and residual fragments produced in $^{238}$U+$^{238}$U at 7.0 MeV/nucleon (logarithmic scale, the black contour lines for primary fragments and colored rectangles for residual fragments). The area of known nuclei are denoted by magenta thick line.}
\label{compare-known-primary-residues}
\end{figure}

The production cross sections for primary fragments produced in reaction $^{238}$U+$^{238}$U at 7.0 MeV/nucleon are calculated by using ImQMD model. In Fig.\ref{compare-known-primary-residues}, the cross sections are plotted by black contour lines. It shows that a large amount of primary fragments are produced via proton and neutron transfer between projectile and target. And the most probable isotopes of primary fragments are located near the line with the isospin asymmetry close to that of $^{238}$U (the isospin asymmetry is 0.227) on the nuclear map. It indicates that most of reaction events have reached the isospin equilibrium at that time. The superheavy primary fragment (114,184) (the isospin asymmetry is 0.235) at the center of the first 'island of stability' denoted by cross symbol in red color is not far from this line.

The production cross sections for residual fragments are obtained through de-excitation of primary fragments by using HIVAP code and shown in Fig.\ref{compare-known-primary-residues} by colored rectangles. Here we set the lower bound cross section to be 10$^{-8}$mb for the production of residual fragments in the figure.
We find that the production cross sections for most of transactinide nuclei are smaller than $10^{-8}$ mb because it is difficult for those primary fragments to survive against fission due to very low fission barrier. For comparison, the area of known nuclei taken from Ref.\cite{Audi1157} are presented by the magenta thick line in the figure. Comparing the predicted produced residual fragments with the known nuclei area, one can find that quite a few unknown neutron-rich isotopes at the 'northeast' area of nuclear map can be produced through multinucleon transfer between two $^{238}$U. Some of those residues are difficult to be produced by fusion reactions. And most of these unknown isotopes are located in the region of actinide elements, and are about three to six neutrons richer than the known most neutron-rich nuclei. For the predicted produced light uranium-like elements with $Z<$ 92, we find that they can reach the border of the proton-rich side of known nuclei in the nuclear map. Because of the high fission barrier, the light uranium-like primary fragments can survive against fission more easily and de-excite through neutron evaporation leading to the production of proton-rich nuclei.

\begin{figure}[hbtp]
\includegraphics[angle=0,scale=0.8]{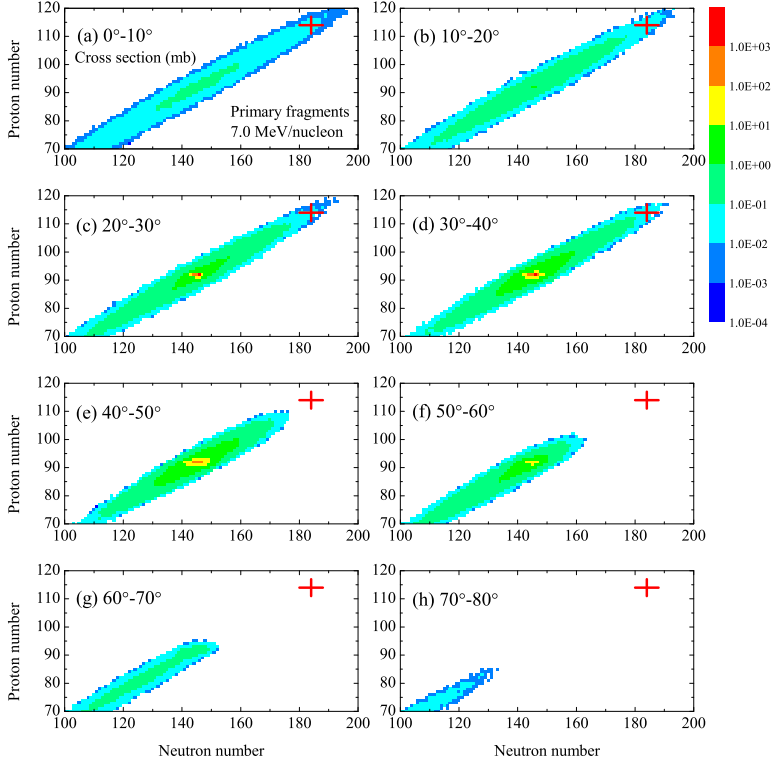}
\caption{(Color online) The landscape of the production cross sections for primary fragments emitted within different laboratory angle range in $^{238}$U+$^{238}$U at 7.0 MeV/nucleon. The red cross symbols denote the center of 'island of stability' ($Z$=114, $N$=184).}
\label{primary-different-angle}
\end{figure}

It is very useful to investigate the outgoing angles of primary and residual fragments for experimental measurement and also for understanding the reaction mechanism. Here we present the calculated results of production cross sections for primary fragments at angle regions of $\theta_{lab}$=0$^\circ$-10$^\circ$,...,70$^\circ$-80$^\circ$ in Fig.\ref{primary-different-angle}. Clearly, the production cross sections for primary fragments vary with their emitting angles and most of transactinide primary fragments are emitted within angles $\theta_{lab}\leq50^\circ$. In the figure, red cross symbols denote the center of 'island of stability' ($Z$=114, $N$=184). It shows that the outgoing angles of primary fragments around (114,184) are within $\theta_{lab}\leq40^\circ$.

\begin{figure}[hbtp]
\includegraphics[angle=0,scale=0.8]{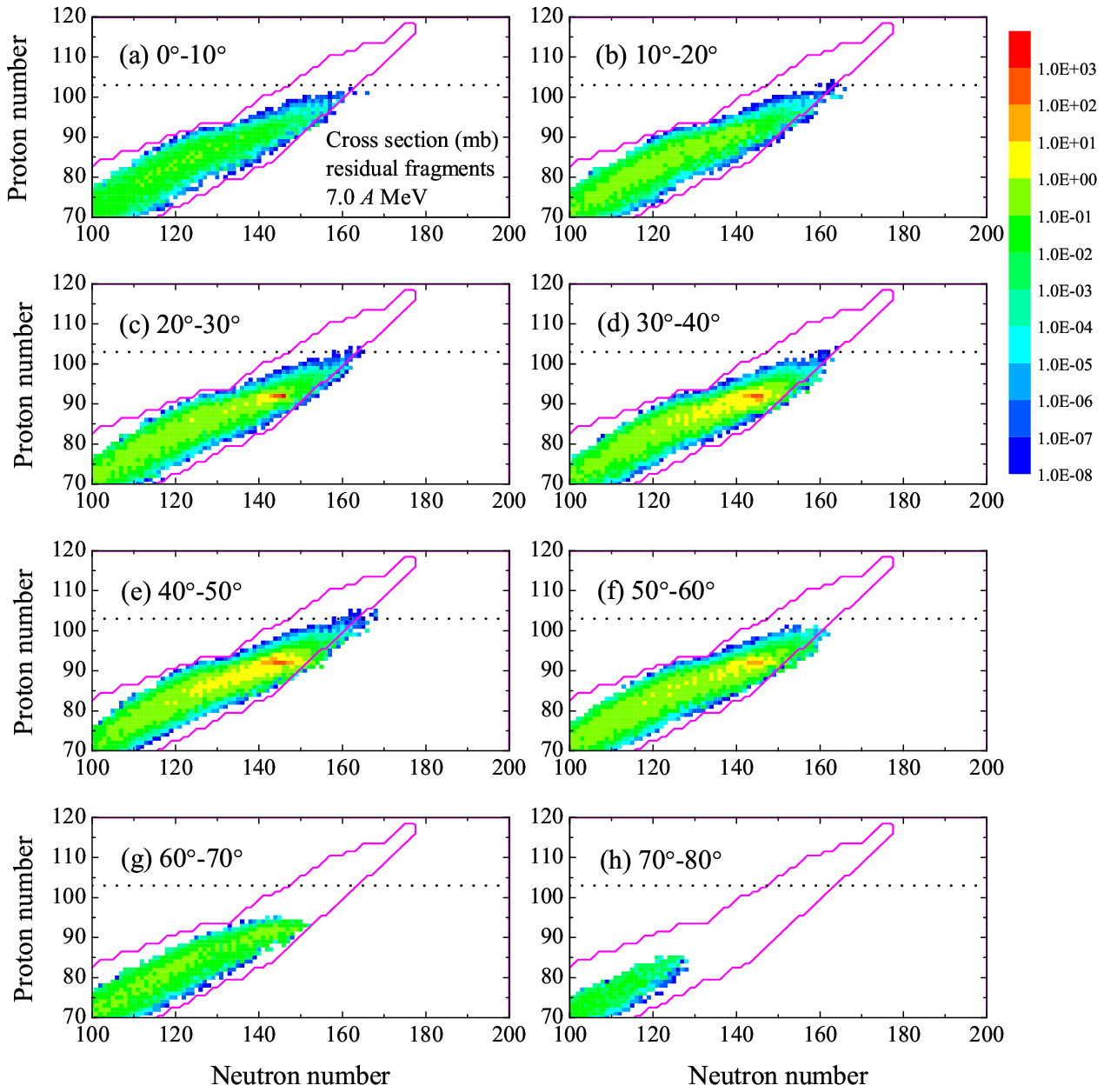}
\caption{(Color online) The landscape of the cross sections for residual fragments emitted within different laboratory angle range in $^{238}$U+$^{238}$U at 7.0 MeV/nucleon. The area of known nuclei are denoted by magenta thick line. The dotted lines denote the position of the heaviest actinide element Lr ($Z$=103).}
\label{residual-angle-known}
\end{figure}

The calculated production cross sections for the residual fragments emitted at angle regions $\theta_{lab}$=0$^\circ$-10$^\circ$,...,70$^\circ$-80$^\circ$ are shown in Fig.\ref{residual-angle-known}. The dotted lines denote the position of the heaviest actinide element Lr ($Z$=103). After the de-excitation process, the production cross sections for superheavy nuclei around the 'island of stability' are smaller than 10$^{-8}$mb.
It is noted from the figure that the outgoing angles of unknown actinide and transactinide isotopes are around $\theta_{lab}$=0$^\circ$-60$^\circ$ and 40$^\circ$-50$^\circ$, respectively.

\begin{figure}[hbtp]
\includegraphics[angle=0,scale=0.6]{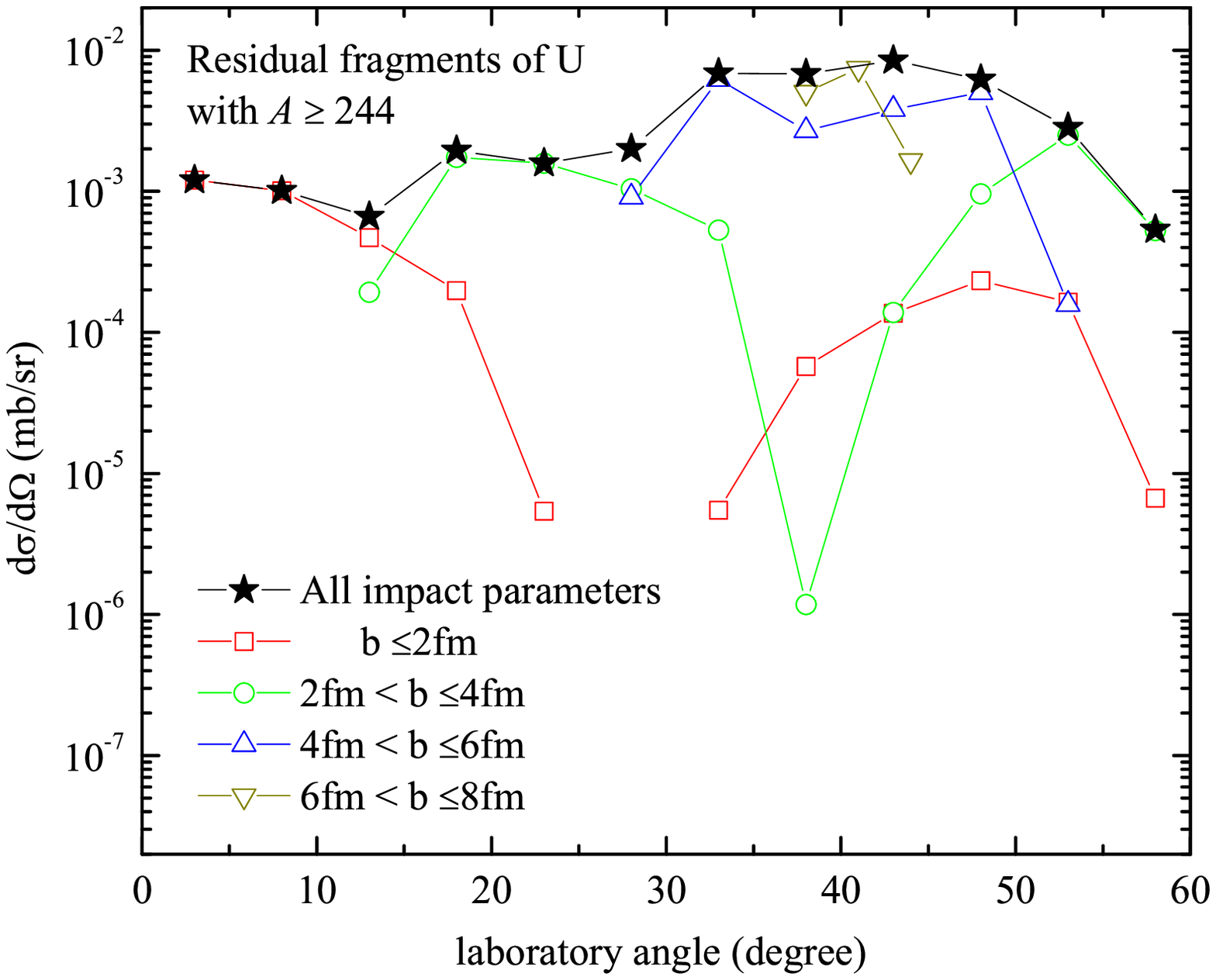}
\caption{(Color online) The angular distribution of unknown isotopes of uranium ($A\geq$ 244) produced in reactions $^{238}$U+$^{238}$U at different impact parameters.}
\label{Unknown-U-angle}
\end{figure}

In order to further investigate the contribution to the production of residual fragments from different outgoing angles shown in Fig.\ref{residual-angle-known} and the production mechanism of unknown neutron-rich nuclei, we take the unknown uranium isotopes with $A\geq$ 244 as an example. The calculated angular distribution of residual fragments of uranium with $A\geq$ 244 is shown in Fig.\ref{Unknown-U-angle}. Here the star symbols denote the total angular distribution of residual fragments of uranium with mass number $A\geq$ 244 and the contributions from the reactions within different impact parameter intervals are shown with the lines with different color and different symbols. For the reactions at the impact parameters $b\leq$ 6 fm, a double-hump distribution is observed, the left hump with smaller outgoing angles corresponds to the residues produced from the target-like primary fragments and the right ones with larger outgoing angles come from the projectile-like primary fragments. With the increasing of impact parameters, the width of the angular distribution for each hump decreases, and the peak of the left hump shifts from an angle of less than 10$^\circ$ to about 32.5$^\circ$, while that of the right hump shifts from about 52.5$^\circ$ to 47.5$^\circ$ and the shift is much smaller compared with that of the left one. The total angular distribution of the residues by adding up the contributions from all impact parameters becomes flat with a wide hump around 30$^\circ$-55$^\circ$, which is the superposition of the contributions from projectile-like and target-like primary fragments. The hump part of the angular distribution mainly comes from the reactions at impact parameters $b$=4-8 fm.

\begin{figure}[hbtp]
\includegraphics[angle=0,scale=0.8]{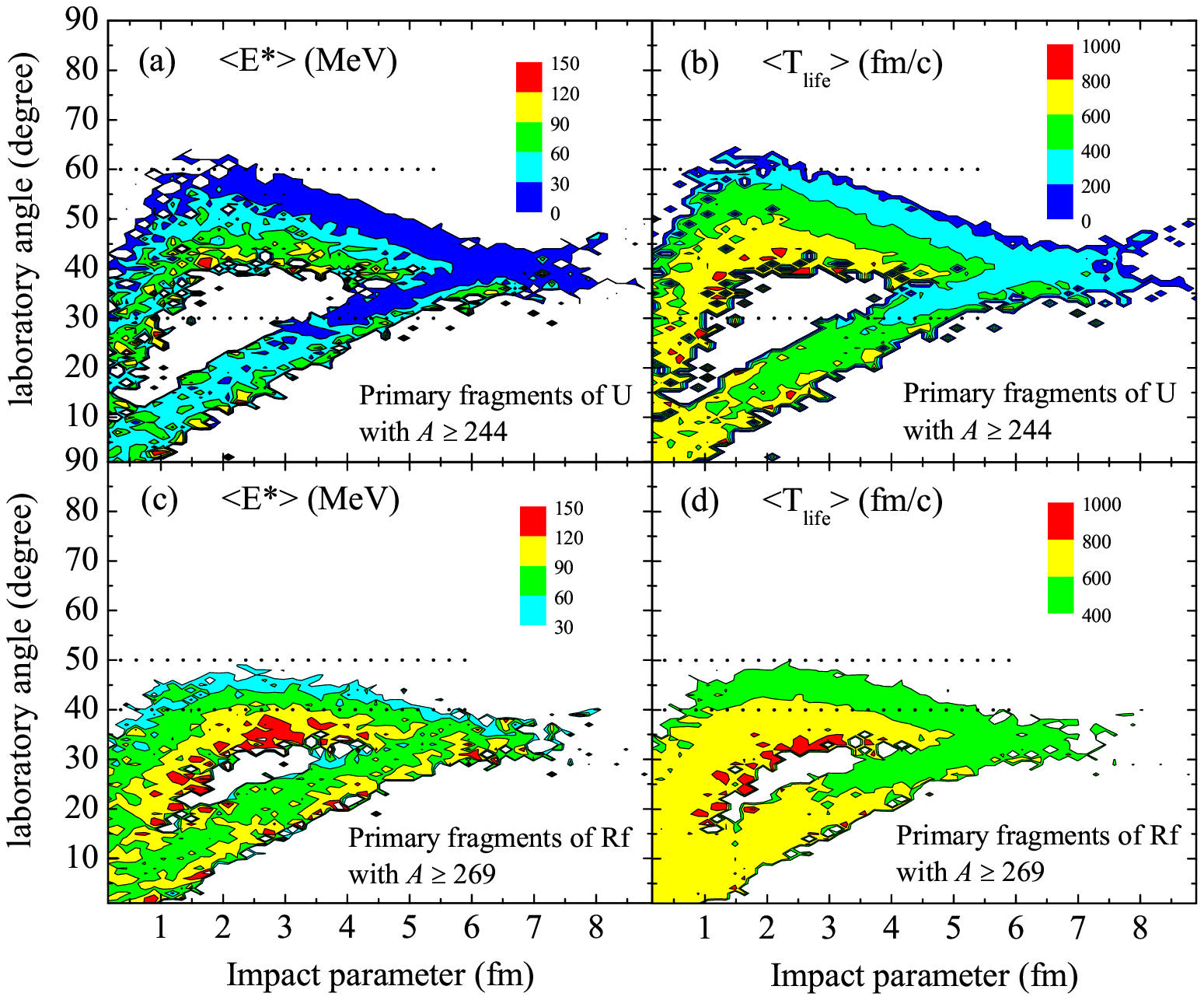}
\caption{(Color online)(a) The average excitation energies of primary fragments of unknown isotopes of uranium and (b) the average lifetime of composite system from which those primary fragments are produced. (c) and (d) are the same as (a) and (b) but for the unknown isotopes of rutherfordium.}
\label{excit-energy-time-U}
\end{figure}

In order to track the origin of the two-hump behavior shown in Fig.\ref{Unknown-U-angle}, we further study the dependence of average excitation energies $<E^{*}>$ of primary fragments of uranium with $A\geq$ 244 and the average lifetime $<T_{life}>$ of composite system formed in the corresponding events on the outgoing angles and impact parameters. In Fig.\ref{excit-energy-time-U},
the $<E^{*}>$ and the $<T_{life}>$ are taken for the primary fragments in a small interval of outgoing angles $\Delta\theta_{lab}$=1$^\circ$
and from the reactions within an impact parameter interval $\Delta{}b$=0.15 fm.
In Fig.\ref{excit-energy-time-U} (a) and (b), one can find that the primary fragments of uranium with $A\geq$ 244 are separated into two branches in both panels: the upper one consists of the projectile-like primary fragments and the lower one consists of target-like primary fragments. And there also exists a correspondence between the excitation energy of primary fragments and the life time of corresponding composite system, i.e. the longer the lifetime of the composite system is, the higher the excitation energy of the primary fragments produced from the composite system is.
Further, the low excitation energy area of $\leq$ 30 MeV (the blue area) in panel (a) coincides with the area with short lifetime area of 200-400fm/c (the light blue area) in panel (b).
Relating Fig.\ref{excit-energy-time-U} (a) and (b) to Fig.\ref{Unknown-U-angle}, we can obtain a scene that most of the residual fragments of uranium with $A\geq$ 244 are produced among the reaction events, in which two uranium nuclei bombarding with impact parameters $b$=4-8 fm contact with each other for about 200-400 fm/c and then the composite system re-separate.

Another example is the production of unknown isotopes of rutherfordium ($Z$=104) with $A\geq$ 269 for investigating the production mechanism of unknown neutron-rich transactinide nuclei.
Fig.\ref{excit-energy-time-U} (c) and (d) show the average excitation energies of primary fragments with $Z=104$ and $A\geq$ 269 and the average lifetimes of their corresponding transient composite systems as the function of impact parameters and outgoing angles.
The same as in panels (a) and (b), in Fig.\ref{excit-energy-time-U} (c) and (d) the primary fragments are also separated into two branches corresponding to projectile-like and target-like primary fragments, respectively.
But if comparing these two panels with (a) and (b) more carefully, we can find large difference between the production of unknown rutherfordium isotopes ($A\geq$ 269) and that of unknown uranium isotopes ($A\geq$ 244). For the Rf case, because of a large number of nucleons (12 protons and over 19 neutrons) being transferred, the collision time between projectile and target become much longer as is seen from panel (d), where the shortest collision time is 400fm/c. Thus the average excitation energies of primary fragments are all larger than 30 MeV.
Moreover the reactions at larger impact parameter have much less contribution compared with the case of the production of unknown uranium isotopes.
Eventually, it leads to a very small cross section for the residual fragments of Rf with $A\geq$ 269.
The outgoing angles of rutherfordium primary fragments are smaller compared with those of uranium primary fragments with $A\geq$ 244 due to the longer collision time of composite system (because of the rotation of the composite system).
From Fig.\ref{residual-angle-known}, one sees that the production cross sections for the residual fragments of Rf with $A\geq$ 269 are lower than 10$^{-6}$mb and the outgoing angles are in a narrow interval of 40$^\circ$-50$^\circ$. From Fig.\ref{residual-angle-known}, Fig.\ref{excit-energy-time-U}(c) and Fig.\ref{excit-energy-time-U}(d), we can deduce that those residues of Rf come from such kind of reaction events, in which the projectile-like fragments capturing a large number of nucleons from target bring a relatively larger collective kinetic energy (i.e. relatively lower excitation energies) and exit with laboratory angles around $\theta_{lab}$=40$^\circ$-50$^\circ$.
From these two examples, we can learn that for the unknown neutron-rich uranium residues, both projectile-like and target-like primary fragments with low excitation energies of $\leq30$ MeV provide the comparable contributions and thus the residual fragments have a wider angular distribution of $\theta_{lab}$=30$^\circ$-60$^\circ$. As the number of transferred protons and neutrons increases, the collision time between projectile and target needed  increases for the corresponding reaction events and the excitation energies of primary fragments become higher. Eventually, the production cross section for the residues of Rf decreases quickly and the outgoing angle of residues becomes narrower. It is because the outgoing angles of residual fragments from projectile-like primary fragments decrease due to the rotation of the composite system.

\begin{center}
\textbf{IV. Summary}
\end{center}

In this work we apply the improved quantum molecular dynamics (ImQMD) model incorporated with the statistical evaporation model (the HIVAP code) to study the reaction $^{238}$U+$^{238}$U at 7.0 MeV/nucleon.  The calculation results of the production cross sections for the primary and residual fragments with charge number from $Z$=70 to 120 are presented. About sixty unknown neutron-rich isotopes from element Ra ($Z$=88) to Db ($Z$=105) with the production cross sections above the lower bound of 10$^{-8}$ mb among the residual fragments produced in the reaction are predicted.
The outgoing angles of primary and residual fragments are also investigated. We find that for most of the unknown neutron-rich isotopes around uranium, the outgoing angles are in a wider range of $\theta_{lab}$=30$^\circ$-60$^\circ$, while for those of heavier transactinide isotopes of Rf the outgoing angles are in a narrower range of 40$^\circ$-50$^\circ$.
In order to understand the production mechanism of unknown neutron-rich isotopes, we study the impact parameter dependence of the excitation energies of primary fragments of uranium isotopes with $A\geq$ 244 and that of rutherfordium isotopes with $A\geq$ 269 and the lifetimes of their corresponding composite systems. We find that for the former case the collision time between two uranium nuclei is shorter and the primary fragments producing those residual fragments have low excitation energies of $\leq30$ MeV and their outgoing angles covers a wider range of 30$^\circ$-60$^\circ$. And for the later case the longer collision time is needed for the transfer of a large number of nucleons and thus it results in the higher excitation energies and smaller outgoing angles of primary fragments and eventually results in a very small production cross sections and a narrower outgoing angle range of 40$^\circ$-50$^\circ$ for the residual fragments of Rf with $A\geq$ 269. This study should be useful for us to select the suitable projectile and target to produce the unknown heavy neutron-rich isotopes.

\begin{center}
\textbf{ACKNOWLEDGMENTS}
\end{center}

This work is supported by the National Natural Science Foundation of China under Grants Nos. (11005155, 11475262, 11275052, 11375062, 11547312, 11475004, 11275068) and National Key Basic Research Development Program of China under Grant No. 2013CB834404. We acknowledge support by the computing server C3S2 in Huzhou University.


\end{document}